\documentclass[amsfonts, amssymb, amsmath, aps, prx, twocolumn, preprintnumbers]{revtex4-1}
\usepackage{graphicx}
\usepackage[pdfencoding=auto, draft=false, colorlinks=true, linkcolor=blue, urlcolor=blue, pdftitle={title}, pdfauthor={Jaeha Lee, Keita Takeuchi, Kaisei Watanabe, and Izumi Tsutsui}]{hyperref}
\usepackage{physics}
\usepackage[sf]{subfigure}

\newcommand{\rvline}{\hspace*{-\arraycolsep}\vline\hspace*{-\arraycolsep}} % vertical lines in block matrices
\begin{document}
\preprint{KEK-TH-2250}

%%%%%%%%%%%%%%%%%%%%%%%%%%%%
\title{Uncertainty Relations of Variances in View of the Weak Value}

\author{Jaeha Lee$^1$}
\email[Email:]{lee@iis.u-tokyo.ac.jp}
\author{Keita Takeuchi$^2$}
\email[Email:]{k-takeuchi@19.alumni.u-tokyo.ac.jp}
\author{Kaisei Watanabe$^2$}
\email[Email:]{wkaisei@post.kek.jp}
\author{Izumi Tsutsui$^3$}
\email[Email:]{izumi.tsutsui@kek.jp}

\affiliation{$^1$Institute of Industrial Science, The University of Tokyo, 5-1-5 Kashiwanoha, Kashiwa, Chiba 277-8574, Japan\\
$^2$Department of Physics, The University of Tokyo, 7-3-1 Hongo, Bunkyo-ku, Tokyo 113-0033, Japan\\
$^3$Theory Center, Institute of Particle and Nuclear Studies,
High Energy Accelerator Research Organization (KEK), 1-1 Oho, Tsukuba, Ibaraki 305-0801, Japan}

\date{\today}
\begin{abstract}
The Schr{\"o}dinger inequality is known to underlie the Kennard-Robertson inequality, which is the standard expression of quantum uncertainty 
for the product of variances of two observables $A$ and $B$, in the sense that the latter is derived from the former.
In this paper we point out that, albeit more subtle, there is yet another inequality which 
underlies the Schr{\"o}dinger inequality in the same sense.  The key component of this observation is the use of the weak-value operator $A_{\rm w}(B)$ introduced in our previous works (named after Aharonov's weak value), which was shown to act as the proxy operator for $A$ when $B$ is measured.  The lower bound of our novel inequality supplements that of the Schr{\"o}dinger inequality by a term representing the discord between $A_{\rm w}(B)$ and $A$.  In addition, the decomposition of the Schr{\"o}dinger inequality, which was also obtained in our previous works by making use the weak-value operator, is examined more closely to analyze its structure and the minimal uncertainty states.  Our results are exemplified with some elementary spin 1 and 3/2 models as well as the familiar case of $A$ and $B$ being the position and momentum of a particle.  
\end{abstract}

\maketitle

%%%%%%%%%%%%%%%%%%%%%%%
\section{Introduction}
\label{sec: intro}
%%%%%%%%%%%%%%%%%%%%%%%

The uncertainty relation epitomizes the indeterminacy of quantum mechanics --- which is a pronouncement made originally by Heisenberg out of his penetrating observations on a number of illuminating instances.  The most notable of them is arguably the Gedankenexperiment of a gamma-ray microscope \cite{Heisenberg}, for which he presented a trade-off relation between error and disturbance associated with the measurement of position $x$ and momentum $p$ of an electron.  His relation, which was gained by heuristic arguments, was soon superseded by a rigorous relation devised by Kennard \cite{Kennard} on the product of variances of $x$ and $p$, which was subsequently generalized by Robertson \cite{Robertson} for arbitrary two observables $A$ and $B$.  The outcome, which we now call the Kennard-Robertson (KR) inequality, has become a standard textbook material to be presented as {\it the} uncertainty relation.  Most notably, it is used for highlighting the quantum nature of physical systems, such as the orders of lowest energy levels of an atom.  Investigation of quantum states which minimizes the uncertainty, {\it i.e.}, the minimum uncertainty states of the KR inequality, leads to the notion of 
coherent states, which have nowadays become a basic tool in studying signal/image processing and quantum optics.  

Whereas the KR inequality gives the restriction on the product of variances of the two observables $A$ and $B$ by a lower bound arising from their noncommutativity expressed in terms of the commutator $[A, B] := AB - BA$, there exists another source of the bound due to the covariance of the two observables expressed in terms of the anticommutator $\{A, B\} := AB + BA$.  These two sources of restriction on the variances are integrated in a single inequality known as the Schr{\"o}dinger inequality \cite{Schroedinger}, which thus furnishes a lower bound tighter than that of the KR inequality.  The states which satisfy the lower bound of the inequality is then called the minimum uncertainty states of the Schr{\"o}dinger inequality (for review articles, see \cite{Angelow2009, Trifonov2001}).

In our earlier works, we found that the Schr{\"o}dinger inequality admits a decomposition into two inequalities according to the two sources, namely quantum indeterminacy and classical covariance \cite{Lee_Tsutsui_1,Lee_Tsutsui_2}.  This was made possible by using our geometrical picture of the operator space utilizing the weak-value operator \cite{Lee_Tsutsui_1,Lee_Tsutsui_2,Lee_Tsutsui_3} named after Aharonov's weak value \cite{ABL,AAV}.  

In this paper, based on our previous results, we further analyze the structure of the Schr{\"o}dinger inequality more extensively.  Our analysis leads to an intriguing finding that the lower bound must be supplemented by an extra term that is proportional to the quantity representing the discord between the two observables $A$ and $B$ with respect to one another \cite{Lee_Tsutsui_1,Lee_Tsutsui_2,Lee_Tsutsui_3}.  We further investigate in detail the circumstances under which the extra term becomes significant and cannot be ignored --- this amounts to finding a novel and tighter inequality for the uncertainty relation --- and demonstrate that there do exist such cases by a couple of elementary examples.  There, we observe that in one of the examples the extra term precisely fills the gap between the variances and the lower bound of the Schr{\"o}dinger inequality.  This suggests that our inequality incorporates a novel aspect of uncertainty which is represented neither by the noncommutativity nor the covariance, but by the said discord of observables.  Various forms of improvement of the Schr{\"o}dinger inequality have been suggested recently \cite{MBP,CWLSQ2019, MP,YXWS2015,SQ2016,HZJ2018}, some of which require optimization processes to establish their lower bounds.  Our new inequality is distinct from all of them in that the improvement is implemented by an explicit term, {\it i.e.}, the aforementioned discord between the two observables, which was shown in our previous works to be closely related to the compatibility of two observables.

The present paper is organized as follows.  We first provide, in Sec.~\ref{sec: preliminaries}, a concise review of the main tools employed in this paper, including our geometrical picture of Hilbert-space operators and the weak-value operator.  In Sec.~\ref{sec: geom_drv}, we revisit our previous results on the decomposition of the Schr{\"o}dinger inequality, and subsequently examine them more closely to analyze how the minimal uncertainty states arise.  We then proceed further in Sec.~\ref{sec: tighter} to argue that the lower bound of the Schr{\"o}dinger inequality can be improved by supplementing a term related to the aforementioned discord of the observables.  Conditions under which the extra term becomes non-vanishing will be discussed, and examples demonstrating such cases are provided in Sec.~\ref{sec: cond}.  The typical continuous case of $A$ and $B$ being position $x$ and momentum $p$ is then mentioned in Sec.~\ref{sec: cont}, and finally, Sec.~\ref{sec: condis} is devoted to the conclusions and discussions.  Two appendices are attached for supplementing the text, one providing derivations of the formulae utilized in the examples, and the other offering a detailed pictorial representation of the various terms concerned in analyzing the Schr{\"o}dinger equality in view of our geometrical picture of operators.

%%%%%%%%%%%%%%%%%%%%%%%
\section{Preliminaries}
\label{sec: preliminaries}
%%%%%%%%%%%%%%%%%%%%%%%

In this section, we first briefly recapitulate our geometric tools and revisit the uncertainty relation of variances in the forms of the Schr{\"o}dinger inequality and the KR inequality.  We then review our definition of the weak-value operator.  In this section, as an introductory warm-up, the weak-value operator will be first presented under a certain set of assumptions that reduces it to a much simpler form than our original general definition; in the next section, some of the assumptions will be gradually lifted to reveal its relatively more general form (for the truly general form defined with full rigour, the reader is referred to our previous works \cite{Lee_Tsutsui_2,Lee_Tsutsui_3}).

Let $\ket{\psi}$ be a quantum state \footnote{%
In the present paper we only consider pure states for simplicity, but the generalization to mixed states is straightforward.
} 
belonging to the Hilbert space ${\cal H}$ of a quantum system, and let ${\cal L}({\cal H})$ be the set of all the linear operators acting on ${\cal H}$.  Given an operator $X \in {\cal L}({\cal H})$, we use the notation $\ev{X}:=\ev{X}{\psi}$ to represent the expectation value associated with the state $\ket{\psi} \in {\cal H}$.  The set ${\cal L}({\cal H})$ is a linear space equipped with the inner product
\begin{align}
    \label{eq: innerprod}
(X,Y) := \ev{X^{\dagger}Y}, \quad \hbox{for}\,\,  X, Y \in {\cal L}({\cal H}),
\end{align}
provided that we identify two operators $X, Y$ for which 
$\norm{X-Y}=0$ holds with the norm defined by $\norm{X}:=\sqrt{(X,X)}$.  
Regarding the complex conjugation, one has
\begin{align}
    \label{eq: compconj}
(X,Y)^* = (Y,X).
\end{align}
Note that the above inner product is {\it state-dependent} and hence has to be considered at every point in the Hilbert space ${\cal H}$ separately.  Nevertheless, it provides us with a geometric picture of the operator space  \cite{Lee_Tsutsui_1,Lee_Tsutsui_2,Lee_Tsutsui_3}, which is a very convenient tool in analyzing the structure of the inequality we are dealing with and even to find a novel inequality tighter than the existing one as we shall see later.

Now, given an observable operator $A$ belonging to the space of self-adjoint operators ${\cal S} \subset {\cal L}({\cal H})$, we introduce the shorthand $\Delta A:=A-\ev{A}$, which allows us to write the statistical variance $\ev{(\Delta A)^2} = \norm{\Delta A}^{2}$ of the observable $A$ in terms of the norm.  The Cauchy-Schwarz (CS) inequality applied to the product of the variances of observables $A, B \in {\cal S}$ then yields
\begin{align}
    \label{eq: CSineq}
    \norm{\Delta A}^{2} \norm{\Delta B}^{2} \geq \abs{(\Delta A, \Delta B)}^{2} = \abs{\ev{\Delta A\, \Delta B}}^{2}.
\end{align}
Based on the identity
\begin{align}
    \Delta A\,\Delta B&=\frac{1}{2}\comm{\Delta A}{\Delta B}+\frac{1}{2}\acomm{\Delta A}{\Delta B} \nonumber \\
    &=\frac{1}{2}\comm{A}{B}+\frac{1}{2}\acomm{\Delta A}{\Delta B},
    \label{eq: DADB}
\end{align}
and further observing that $\ev{\comm{A}{B}}$ is purely imaginary whereas $\ev{\acomm{\Delta A}{\Delta B}} = \ev{\acomm{A}{B}}-2\ev{A}\!\ev{B}$ is real, one finds that the CS inequality \eqref{eq: CSineq} is equivalent to 
\begin{align}
    \norm{\Delta A}^{2} \norm{\Delta B}^{2}
    \geq\abs{\frac{1}{2}\ev{\comm{A}{B}}}^{2}\!+\abs{\frac{1}{2}\ev{\acomm{A}{B}}-\ev{A}\!\ev{B}}^{2},
    \label{eq: Schr}
\end{align}
which is nothing but the Schr{\"o}dinger inequality \cite{Schroedinger}.  The KR inequality \cite{Kennard, Robertson}
\begin{align}\label{eq: KR}
    \norm{ \Delta A} \,\norm{ \Delta B} \geq \frac{1}{2}\abs{\ev{\comm{A}{B}}}
\end{align}
then follows trivially from \eqref{eq: Schr}.

Since the Schr{\"o}dinger inequality \eqref{eq: Schr} is essentially the CS inequality, its lower bound is attained if and only if (iff)
\begin{align} 
\label{eq: hold_Schr}
    \Delta A\ket{\psi}=(\lambda+i\mu)\,\Delta B\ket{\psi}
\end{align}
for some real numbers $\lambda, \mu$.  The states which satisfy the condition \eqref{eq: hold_Schr} are called the minimum uncertainty states of the Schr{\"o}dinger inequality.

Next, we review our weak-value operator named after Aharonov's weak value.  The weak value offers a novel measurable quantity that potentially represents the physical value of $A$ when the system is measured weakly under a prescribed quantum process of transition.  The process is specified by the pair of an initial state $\ket{\psi}$ and a final state $\ket{\phi}$.  They are occasionally called the \lq preselected state\rq\, and the \lq postselected state\rq, respectively, to emphasize the fact that both states can be freely chosen by the observer.  In other words, $\ket{\phi}$ is not defined as the state determined uniquely from $\ket{\psi}$ under the unitary time evolution described by the Hamiltonian of the system.  The \emph{weak value} \cite{ABL,AAV} is then defined as 
\begin{align}\label{def:weak_value}
A_{\rm w} := \frac{\mel{\phi}{A}{\psi}}{\ip{\phi}{\psi}}
\end{align}
from the preselected state and the postselected state that we have chosen, provided that the amplitude of the quantum transition between them is non-vanishing ${\ip{\phi}{\psi}} \ne 0$.

For our purpose, we would like to choose one of the eigenstates $\ket{b_{i}}$ of an observable $B$ as the postselected state $\ket{\phi}$, where $i$ labels each of its distinct eigenvalues, {\it i.e.}, $B \ket{b_{i}} = b_i \ket{b_{i}}$.  We here temporarily assume for simplicity of arguemnts that (i) ${\cal H}$ is finite dimensional, (ii) $B$ is non-degenerate, and further that (iii)  the transition process is not void, {\it i.e.}, $\ip{b_{i}}{\psi}\neq 0$.  Under the assumptions (i)--(iii), the \emph{weak-value function} \cite{Lee_Tsutsui_1,Lee_Tsutsui_2,Lee_Tsutsui_3} introduced in our previous works reduces to the simple form
\begin{align}
    A_{\rm w}(b_{i}) := \frac{\mel{b_{i}}{A}{\psi}}{\ip{b_{i}}{\psi}},
    \label{eq: def_dscrt_wv}
\end{align}
which is nothing but the weak value $A_{\rm w} = A_{\rm w}(b_{i})$ with the choice $\ket{\phi} = \ket{b_{i}}$.  Under the same assumptions, the \emph{weak-value operator} \cite{Lee_Tsutsui_1, Lee_Tsutsui_2, Lee_Tsutsui_3} reduces to
\begin{align}
\label{eq: def_wvop}
    A_{\rm w}(B):=\sum_{i}A_{\rm w}(b_{i})\op{b_{i}},
\end{align}
where the summation is taken over all the eigenvalues of the observable $B$. Note that $A_{\rm w}(b_{i})$ may not be real and, accordingly, 
the weak-value operator $A_{\rm w}(B)$ may not be self-adjoint in general.

In fact, the above weak-value operator is just a special case of the operators constructed from 
a self-adjoint operator admitting the spectral decomposition $B = \sum_i  b_i\, \Pi_i$ and 
a complex-valued function $f(b_i)$ as
\begin{align}
\label{eq: def_fopgenb}
f(B):=\sum_i  f(b_i)\, \Pi_i,
\end{align}  
where $\Pi_i$ is the projection operator onto the eigenspace associated with the eigenvalue $b_i$.  Under our current assumptions in which $B$ is free from degeneracy, we have $\Pi_i = \op{b_{i}}$ as a special case.

As we have mentioned in the beginning of this section, the above definition of the weak-value operator \eqref{eq: def_wvop}, as well as that of the general operators \eqref{eq: def_fopgenb}, admit generalization even to those $B$ that possess continuous spectra, in which case the summation appearing in the above definitions should be formally \lq replaced\rq\ by the integral over the spectrum (see \cite{Lee_Tsutsui_2, Lee_Tsutsui_3} for their general and rigorous definitions and constructions).

Now, such operators $f(B)$ generated by $B$ form a linear subspace ${\cal L}_B \subset {\cal L}({\cal H})$ 
which is closed under conjugation, that is,
\begin{align}
f^\dagger(B) =\sum_i  f^*(b_i)\,\Pi_i \in {\cal L}_B
\end{align}  
for any $f(B) \in {\cal L}_B$.  
Clearly, multiplication of operators in the space ${\cal L}_B$ is commutative,  
\begin{align}
\label{eq: comop}
 f(B) \,g(B) = g(B)\, f(B)
\end{align}
for $f(B), \, g(B) \in {\cal L}_B$.
Although the operator $f(B)$ is not self-adjoint unless $f$ is a real-valued function, one may consider its real and imaginary part
 \begin{align}
\label{eq: rimdecomp}
 f(B) = \Re f(B) + i \Im f(B)
 \end{align}
in terms of the components
 \begin{align}
 \label{reimdef}
        \Re f(B)&:=\frac{f(B)+f^{\dagger}(B)}{2},  \nonumber \\
        \Im f(B)&:=\frac{f(B)-f^{\dagger}(B)}{2i},
\end{align}
which are both self-adjoint.   
The norms of these operators fulfill the identity
\begin{align}
\label{comsum}
    \norm{f(B)}^2=  \norm{\Re f(B)}^2 +  \norm{\Im f(B)}^2,
\end{align}
valid for any $f(B) \in {\cal L}_B$.
The properties \eqref{eq: comop} and \eqref{comsum} suggest that one may treat the operators $ f(B)$ as if they were complex numbers, as long as one is working with the linear space ${\cal L}_B$.

Returning to the finite dimensional case where the set of assumptions (i)--(iii) are fulfilled, we refer to a remarkable identity \cite{Lee_Tsutsui_1} regarding the weak-value operator:
\begin{align}
A_{\rm w}(B)\ket{\psi} =\sum_{i}\frac{\mel{b_{i}}{A}{\psi}}{\ip{b_{i}}{\psi}}\ket{b_{i}}\ip{b_{i}}{\psi} = A\ket{\psi}.
    \label{eq: AwBpsi}
\end{align}
This is to say that the action of the weak-value operator \eqref{eq: def_wvop} on the preselected state coincides with that of the original operator $A$.  This property follows immediately from the completeness condition $\sum_{i} \ket{b_{i}}\bra{b_{i}} = I$, where $I$ is the identity operator in ${\cal L}({\cal H})$.
The equality \eqref{eq: AwBpsi} may at first sound strange, because the r.h.s.\ is independent of $B$ while the l.h.s.\ contains the weak-value operator $A_{\rm w}(B)$ which is $B$-dependent.  However, the $B$-dependence disappears under the particular combination of the operator $A$ with the state $\ket{\psi}$ in the form of the weak-value operator $A_{\rm w}(B)$; in other words, $A_{\rm w}(B)$ is designed to reproduce the result of the action of $A$ on the state $\ket{\psi}$ by means of an operator generated by $B$, even if $B$ does not necessarily commute with $A$.  This shows that the weak-value operator $A_{\rm w}(B)$ acts as a proxy operator of optimal choice for approximation of observables and estimation of physical parameters \cite{Lee_Tsutsui_1,Lee_Tsutsui_2}.

%%%%%%%%%%%%%%%%%%%%%%%
\section{Decompsing the Schr{\"o}dinger Inequality}
\label{sec: geom_drv}
%%%%%%%%%%%%%%%%%%%%%%%

Armed with the necessary tools to analyze the structure of the operator space ${\cal L}({\cal H})$ with respect to a given state $\ket{\psi}$, we next revisit the decomposition of the Schr{\"o}dinger inequality obtained in our prior works \cite{Lee_Tsutsui_1,Lee_Tsutsui_2}; although the result has been originally demonstrated for the general case, in this section, we will recapitulate it under the assumptions (i)--(iii) for the sake of demonstration.  Based on this decomposition, we further point to the minimum uncertainty states of the inequality according to the components that appear in the decomposition separately.

\begin{figure}[t]
    \includegraphics[scale=0.25]{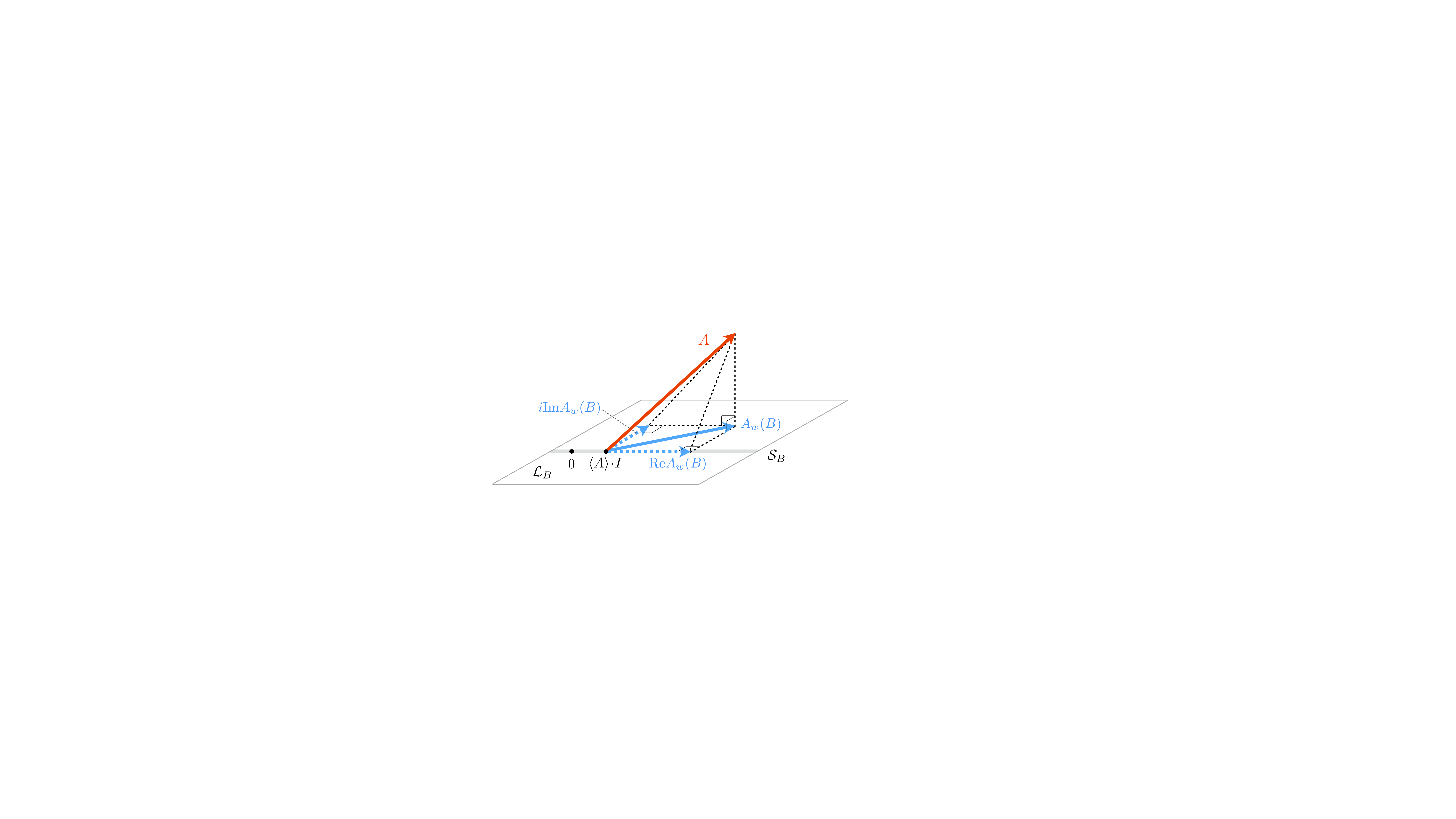}
    \caption{Geometric relations among the operators involved \cite{Lee_Tsutsui_1,Lee_Tsutsui_2}. The projection of the operator $A$ onto ${\cal L}_B$ is its weak-value operator $A_{\rm w}(B)$. As $A_{\rm w}(B)$ decomposes orthogonally into the Hermitian operator $\Re A_{\rm w}(B)$ and the anti-Hermitian operator $i\Im A_{\rm w}(B)$, the Schr{\"o}dinger inequality decomposes into the two inequalities on each of them as shown in Fig.~\ref{fig: ineq_geom}. The right angle symbol indicates the validity of the \lq Pythagorean identity\rq\ for the sides of the triangle with the symbol.}
    \label{fig: operator}
\end{figure}

We start our review by observing that the identity \eqref{eq: AwBpsi} allows us to rewrite the variance of $A$ as 
\begin{align}
\norm{\Delta A}^{2} = \norm{A-\ev{A}}^{2} = \norm{A_{\rm w}(B) -\ev{A}}^{2}.
\label{predadecop}
\end{align}
Since $A_{\rm w}(B) -\ev{A} \in {\cal L}_B$, one may choose this as $f(B)$ in \eqref{comsum} to obtain
\begin{align}
\label{dadecop}
\norm{A_{\rm w}(B) -\ev{A}}^{2} = \norm{\Re A_{\rm w}(B) -\ev{A}}^{2} + \norm{\Im A_{\rm w}(B)}^{2}.
\end{align}
Secondly, we note that, with the inner product introduced in \eqref{eq: innerprod}, the identity \eqref{eq: AwBpsi} implies
\begin{align}
    (A, f(B))=(A_{\rm w}(B), f(B))
    \label{eq: isp_proj}
\end{align}
for any $f(B) \in {\cal L}_B$.
This indicates that the projection of $A$ onto the subspace ${\cal L}_B$ is given by $A_{\rm w}(B)$ (see Fig.~\ref{fig: operator}).
In particular, when $f(B) \in {\cal S}_B \subset {\cal L}_B$ is self-adjoint, where ${\cal S}_B$ denotes the space of self-adjoint operators generated by $B$ (which consists of those operators that are constructed from real-valued functions), one has
\begin{align}
\label{twoimpidentf}
\Re(A, f(B)) %&= \Re(A_{\rm w}(B), f(B)) 
&=\hphantom{-}(\Re A_{\rm w}(B), f(B)), \\ 
\Im(A, f(B)) %&= \Im(A_{\rm w}(B), f(B)) 
&=-(\Im A_{\rm w}(B), f(B)),
\label{twoimpidents}
\end{align}
which can be readily confirmed from \eqref{reimdef} and \eqref{eq: isp_proj}.  

In order to form the product of variances, we multiply both sides of \eqref{dadecop} by $\norm{\Delta B}^{2}$, and subsequently apply the CS inequality to the product appearing in the first term in the r.h.s.\ to obtain
\begin{align}
\norm{\Re A_{\rm w}(B) -\ev{A}}^{2}\norm{\Delta B}^{2} 
&\geq \abs{(\Re A_{\rm w}(B) -\ev{A}, \Delta B)}^2 \nonumber \\
&= \abs{(\Re A_{\rm w}(B), \Delta B)}^2 \nonumber \\
&= \abs{\Re (A, \Delta B)}^2 \nonumber \\
&= \abs{\Re (A, B) - \ev{A}\ev{B}}^2, 
\end{align}
where \eqref{twoimpidentf} is used.
Then, since $\Re (A, B) = \Re\ev{AB} = \frac{1}{2}\ev{\acomm{A}{B}}$, one finds \cite{Lee_Tsutsui_1,Lee_Tsutsui_2}
\begin{align}
\norm{\Re A_{\rm w}(B)-\expval{A}} \,\norm{\Delta B}&\geq \abs{\frac{1}{2}\expval{\acomm{A}{B}}-\expval{A}\expval{B}}.
\label{eq: cov}
\end{align}
This gives one half of the lower bound of the Schr{\"o}dinger inequality \eqref{eq: Schr}.  

The remaining half will come from the product appearing in the second term in the r.h.s.\ of \eqref{dadecop} after $\norm{\Delta B}^{2}$ is multiplied.  In a similar vein, by applying the CS inequality, one obtains
\begin{align}
\norm{\Im A_{\rm w}(B)}^{2}\norm{\Delta B}^{2} 
&\geq \abs{(\Im A_{\rm w}(B), \Delta B)}^2 \nonumber \\
&= \abs{\Im (A, \Delta B)}^2 \nonumber \\
&= \abs{\Im (A, B)}^2,
\end{align}
where we have used \eqref{twoimpidents}.
From $\Im (A, B) = \Im\ev{AB} = \frac{1}{2}\ev{\comm{A}{B}}$, one finds \cite{Lee_Tsutsui_1,Lee_Tsutsui_2}
\begin{align}
\norm{\Im A_{\rm w}(B)} \,\norm{\Delta B}&\geq \abs{\frac{1}{2}\expval{\comm{A}{B}}}, 
\label{eq: KRlike}
\end{align}
which is the remaining half of the Schr{\"o}dinger inequality \eqref{eq: Schr} as expected. 

In summary, we have decomposed the variance $\norm{\Delta A}^{2}$ into two terms, one containing 
$\Re A_{\rm w}(B)$ and the other $\Im A_{\rm w}(B)$, and then applied the CS inequality separately for the two terms after the variance 
$\norm{\Delta B}^2$ is multiplied. The result is \eqref{eq: cov} and  \eqref{eq: KRlike}, which are obviously the two components giving the lower bound of the Schr{\"o}dinger inequality \eqref{eq: Schr}.  This decomposed set of inequalities is, in principle, more informative than the combined one; since the latter form \eqref{eq: Schr} can be derived from the former by summing the squares of the both-hand sides of \eqref{eq: cov} and \eqref{eq: KRlike}, and applying \eqref{dadecop} to the l.h.s., but the converse is not always true. 

Since both \eqref{eq: cov} and \eqref{eq: KRlike} are based on the CS inequality, their lower bounds hold iff
\begin{align}
    \label{eq: hold_cov}
    (\Re A_{\rm w}(B)-\expval{A})\ket{\psi} &=\lambda \,\Delta B\ket{\psi}, \\
    \label{eq: hold_KRlike}
    \Im A_{\rm w}(B)\ket{\psi} &=\mu \,\Delta B\ket{\psi},
\end{align}
for some real numbers $\lambda, \,\mu$, respectively. The lower bound of the Schr{\"o}dinger inequality \eqref{eq: Schr} holds when both are satisfied simultaneously.  Indeed, by using \eqref{eq: rimdecomp} and \eqref{eq: AwBpsi}, and combining with \eqref{eq: hold_cov} and \eqref{eq: hold_KRlike} we find
\begin{align}
    \Delta A\ket{\psi} 
    &= \left(A_{\rm w}(B)-\expval{A}\right) \ket{\psi} \nonumber \\
    &= \left[ (\Re A_{\rm w}(B)-\expval{A}) + i\Im A_{\rm w}(B)\right] \ket{\psi} \nonumber \\
    &= (\lambda+i\mu)\Delta B\ket{\psi}, 
\end{align}
which is precisely the minimum uncertainty condition \eqref{eq: hold_Schr} for the Schr{\"o}dinger inequality. In Sec.~\ref{sec: cont}, we shall discuss how our decomposed conditions \eqref{eq: hold_cov} and \eqref{eq: hold_KRlike} work when we obtain the minimum uncertainty states for 
the familiar case of position and momentum.

%%%%%%%%%%%%%%%%%%%%%%%
\section{Tighter Inequalities}
\label{sec: tighter}
%%%%%%%%%%%%%%%%%%%%%%%

Based on our geometric description of operators, the weak-value operator, and the decomposition of the Schr{\"o}dinger inequality, which we have just reviewed in Secs.~\ref{sec: preliminaries} and \ref{sec: geom_drv}, we now show that it is possible to improve the Schr{\"o}dinger inequality into the new inequality
\begin{align}
    \norm{\Delta A}^{2} \norm{\Delta B}^{2} \geq 
    &\abs{\frac{1}{2}\ev{\comm{A}{B}}}^{2}+\abs{\frac{1}{2}\ev{\acomm{A}{B}}-\ev{A}\ev{B}}^{2}\nonumber\\
    &+E(A, B),
    \label{eq: tighter}
\end{align}
where 
\begin{align}
    E(A, B) := \norm{A-A_{\rm w}(B)}^{2} \norm{\Delta B}^{2},
    \label{eq: lastterm}
\end{align}
is an extra term that supplements the lower bound.  Clearly, this inequality becomes strictly tighter than the Schr{\"o}dinger inequality \eqref{eq: Schr} as long as the 
extra term $E(A, B)$ is non-vanishing.  

Before we proceed further, we shall remark that, once the inequality \eqref{eq: tighter} is established, one can always exchange $A$ and $B$ for one another to have $E(B, A)$ in place of $E(A, B)$ in the above inequality.  This suggests that one may also obtain an inequality that is symmetric with respect to the two observable $A$ and $B$.  For instance, one may trivially replace $E(A, B)$ in \eqref{eq: tighter} with the maximum
\begin{align}
    E_{\text{max}}(A, B):=\max\qty{E(A, B),\ E(B, A)}
    \label{eq: tighter_symm}
\end{align}
of the two asymmetric terms, and the inequality still remains valid; in fact, this actually offers a better lower bound than either of those realized by the asymmetric terms $E(A, B)$ or $E(B, A)$.
Another way to obtain a symmetric inequality is to replace $E(A, B)$ with
\begin{align}
    \widetilde{E}(A, B):=\norm{A-A_{\rm w}(B)}^{2}\norm{B-B_{\rm w}(A)}^{2},
    \label{eq: tighter_indp}
\end{align}
which is possible in view of $\norm{\Delta B} \ge \norm{B-B_{\rm w}(A)}$, although this offers the worst lower bound among all the other possible inequalities we have mentioned so far.  Still, the choice \eqref{eq: tighter_indp} has an advantage in that the extra term consists purely of the contributions of the degree of proximity \cite{Lee_Tsutsui_1,Lee_Tsutsui_2,Lee_Tsutsui_3} between the two observables $A$ and $B$ in terms of the other.  In this paper, we solely focus on the inequality \eqref{eq: tighter} that adopts the asymmetric term $E(A, B)$ in order to simplify our discussion; however, the following line of analysis should be, in principle, applicable to the other forms $E_{\text{max}}(A, B)$ and $\widetilde{E}(A, B)$ as well.

Now we shall discuss when our extra term $E(A, B)$ is non-vanishing, which amounts to proving the inequality \eqref{eq: tighter} as well.  For this, we first note that the key property \eqref{eq: AwBpsi} appears to suggest that the discord between the operators $A$ and $A_{\rm w}(B)$ vanishes identically
\begin{align}
    \norm{A-A_{\rm w}(B)}=0,
    \label{eq: ftzero}    
\end{align}
and hence $E(A, B) = 0$.  

At this point, we must point to the fact that the property \eqref{eq: AwBpsi} has only been shown to be valid under the introductory conditions (i)--(iii) we have been assuming so far.  For the purpose of this paper, it is necessary to lift the assumptions (ii) and (iii), which amounts to include the case in which $B$ is degenerate or $\ip{b_{i}}{\psi} = 0$ for some $i$, leaving us with the finite dimensional condition (i).

Under such condition, our weak-value function \cite{Lee_Tsutsui_1,Lee_Tsutsui_2,Lee_Tsutsui_3} is shown to reduce to
\begin{align}
    A_{\rm w}(b_{i}):=\begin{cases}
        \displaystyle\frac{\ev{\Pi_{i}A}}{\ev{\Pi_{i}}}
        & \hbox{for}\, \ev{\Pi_{i}}\neq 0,\\
        c_{i}, & \hbox{for}\, \ev{\Pi_{i}}=0,
    \end{cases}\label{eq: wvdef_gen}
\end{align}
where $\Pi_{i}$ are the projection operators introduced in \eqref{eq: def_fopgenb} and $c_{i}$ are arbitrary complex numbers.  The arbitrariness of $c_{i}$ does not affect the resultant inequality \eqref{eq: tighter} as we shall confirm by means of direct computation in \eqref{eq: norm_rep} later.  We note again that the last assumption (i) can be also lifted so that continuous observables --- such as the position and momentum operators discussed in Sec.~\ref{sec: cont} --- may receive proper treatment (see our previous works \cite{Lee_Tsutsui_2,Lee_Tsutsui_3} for the general and rigorous formulation); in simple terms, this will be formally achieved by replacing the summation by integration in \eqref{eq: wvdef_gen}.

An important fact is that, while \eqref{eq: AwBpsi} becomes no longer valid in general once we lift the assumptions (ii)--(iii), the identity \eqref{eq: isp_proj} still remains valid even for the most general case \cite{Lee_Tsutsui_1,Lee_Tsutsui_2,Lee_Tsutsui_3}.  To confirm this under the assumption (i), we shall resort to  direct computation in this paper.  To this end, we first note that, based on the expression \eqref{eq: wvdef_gen}, the weak-value operator reduces to
\begin{align}
    A_{\rm w}(B) =\sum_{\substack{i\\\ev{\Pi_{i}}\neq0}}\frac{\ev{\Pi_{i}A}}{\ev{\Pi_{i}}}\Pi_{i} + \sum_{\substack{i\\\ev{\Pi_{i}}=0}}c_{i}\Pi_{i}.
    \label{eq: gen_AwB}
\end{align}
Since $\ev{\Pi_{i}}=0$ implies $\Pi_{i}\ket{\psi}=0$, we find that 
the action of $A_{\rm w}(B)$ on the state $\ket{\psi}$ reads
\begin{align}
    A_{\rm w}(B)\ket{\psi} =\sum_{\substack{i\\\ev{\Pi_{i}}\neq0}}\frac{\ev{\Pi_{i}A}}{\ev{\Pi_{i}}}\Pi_{i}\ket{\psi}, 
    \label{eq: actgen_AwB}
\end{align}
which may not be equal to $A\ket{\psi}$, thereby potentially invalidating the identity \eqref{eq: AwBpsi} as mentioned.  On the other hand, the validity of \eqref{eq: isp_proj} can be directly confirmed as
\begin{align}
    (A_{\rm w}(B), f(B))&=\ev{\sum_{\substack{i\\\ev{\Pi_{i}}\neq0}}\frac{\ev{\Pi_{i}A}^*}{\ev{\Pi_{i}}}\Pi_{i}\sum_{j}f(b_{j})\Pi_{j}} \nonumber\\
    &=\ev{\sum_{\substack{i\\\ev{\Pi_{i}}\neq0}}\frac{\ev{\Pi_{i}A}^*}{\ev{\Pi_{i}}}\Pi_{i}f(b_{i})} \nonumber\\
    &=\sum_{\substack{i\\\ev{\Pi_{i}}\neq0}}\ev{\Pi_{i}A}^*f(b_{i}) \nonumber\\
    &=\ev{A\sum_{\substack{i\\\ev{\Pi_{i}}\neq0}}   f(b_{i})\Pi_{i}} \nonumber\\
    &=\ev{A\qty\Bigg(\sum_{\substack{i\\\ev{\Pi_{i}}\neq0}}f(b_{i})\Pi_{i}+\sum_{\substack{i\\\ev{\Pi_{i}}=0}}f(b_{i})\Pi_{i})} \nonumber\\
    &=(A, f(B))
    \label{eq: ev_hBA}
\end{align}
by making use of the identity $\Pi_i\Pi_j = \delta_{ij}\Pi_i$.

In what follows, we make use of the identity \eqref{eq: isp_proj} to point to our desired inequality \eqref{eq: tighter}.  For this, we just choose $f(B) = A_{\rm w}(B) -\ev{A}$ in \eqref{eq: isp_proj} to obtain
\begin{align}
    (A -  A_{\rm w}(B), A_{\rm w}(B)-\ev{A}) = 0.
\end{align}
This allows us to exploit the \lq Pythagorean identity\rq\ (see Fig.~\ref{fig: operator})
\begin{align}
    \norm{\Delta A}^{2} =\norm{A-A_{\rm w}(B)}^{2}+\norm{A_{\rm w}(B)-\ev{A}}^{2}
    \label{eq: da_pyth}
\end{align}
to arrive at our inequality \eqref{eq: tighter} immediately by following the same procedure for obtaining the decomposition \eqref{eq: cov} and \eqref{eq: KRlike} of the Schr{\" o}dinger inequality, starting from
\eqref{dadecop} applied to the second term in the r.h.s.\ of \eqref{eq: da_pyth}.   

With respect to the decomposed form, we can argue that it still leads to the original form \eqref{eq: Schr} in the same way we have shown in Sec.~\ref{sec: geom_drv}, using \eqref{eq: da_pyth} additionally.  We also notice that it can attain the minimum uncertainty even when $E(A, B)\neq0$, where the original form cannot attain as is evident in the new inequality \eqref{eq: tighter}.

We have reviewed in Sec.~\ref{sec: geom_drv} that, under the elementary assumptions (i)--(iii), the term \eqref{eq: ftzero} trivially vanishes, thereby reducing the general form \eqref{eq: da_pyth} to the special case \eqref{predadecop}.  This implies that, in order for our inequality \eqref{eq: tighter} to be truly different from the Schr{\" o}dinger inequality and becomes nontrivial $E(A, B) \ne 0$, we must examine the cases where (ii) and (iii) are no longer assumed.   In the next section, we shall analyze our inequality more closely and provide some elementary examples where $E(A, B) \ne 0$ is indeed realized.

%%%%%%%%%%%%%%%%%%%%%%%
\section{properties of the extra terms}
\label{sec: cond}
%%%%%%%%%%%%%%%%%%%%%%%

In this section, we explore our inequality \eqref{eq: tighter} and observe 
two important properties on the extra term $E(A, B)$, before presenting explicit examples for which $E(A, B) \ne 0$.  
The first property is that (a) if the dimension of the Hilbert space ${\cal H}$ is less than three, then
we have $E(A, B) = 0$ identically and hence our inequality \eqref{eq: tighter} reduces to the Schr{\"o}dinger inequality \eqref{eq: Schr}. 
In other words, for our inequality \eqref{eq: tighter} to be significant, we need $\dim\mathcal{H} \ge 3$.
The second is that (b) if the operator $B$ has only two distinct eigenvalues, then the lower bound of our inequality is always attained irrespective of the choice of the state $\ket{\psi}$.  

Both of these properties require us to consider the general case where $E(A, B) \ne 0$, and to this end
we first choose $f(B)=A_{\rm w}(B)$ in the identity \eqref{eq: isp_proj} to obtain
\begin{align}
(A, A_{\rm w}(B)) = (A_{\rm w}(B), A) = \norm{A_{\rm w}(B)}^{2},
\end{align}
which allows us to rewrite (see Fig.~\ref{fig: operator} for the geometric interpretation \cite{Lee_Tsutsui_1,Lee_Tsutsui_2,Lee_Tsutsui_3})
\begin{align}
    \norm{A-A_{\rm w}(B)}^{2}=\norm{A}^{2}-\norm{A_{\rm w}(B)}^{2}.
    \label{eq: norm_subt}
\end{align}
Our concern is the second term in the r.h.s.\ of \eqref{eq: norm_subt}, which upon inserting \eqref{eq: actgen_AwB} becomes
\begin{align}
    \norm{A_{\rm w}(B)}^{2}&=\ev{\sum_{i}\abs{A_{\rm w}(b_{i})}^{2}\Pi_{i}} \nonumber\\
    &=\sum_{\substack{i\\\ev{\Pi_{i}}\neq0}}\abs{\frac{\ev{\Pi_{i}A}}{\ev{\Pi_{i}}}}^{2}\ev{\Pi_{i}}+\sum_{\substack{i\\\ev{\Pi_{i}}=0}}\abs{c_{i}}^{2}\ev{\Pi_{i}} \nonumber\\
    &=\sum_{\substack{i\\\ev{\Pi_{i}}\neq0}}\ev{A\frac{\op{\Pi_{i}\psi}}{\ev{\Pi_{i}}}A}{\psi}. 
    \label{eq: wvonorm}
\end{align}
Since for non-degenerate $i$ for which $\Pi_{i}=\op{b_{i}}$, the summand in \eqref{eq: wvonorm} is simplified as
\begin{align}
    \frac{\op{\Pi_{i}\psi}}{\ev{\Pi_{i}}}=\frac{\ket{b_{i}}\ip{b_{i}}{\psi}\ip{\psi}{b_{i}}\bra{b_{i}}}{\ip{\psi}{b_{i}}\ip{b_{i}}{\psi}}=\op{b_{i}}=\Pi_{i},
\end{align}
which implies that components for non-degenerate $i$ in the second term 
cancels with those in the first term
$\norm{A}^{2}=\sum_{i}\ev{A\Pi_{i}A}$
in the r.h.s.\ of \eqref{eq: norm_subt}, leaving us with
\begin{align}
    \norm{A-A_{\rm w}(B)}^{2}
    =\sum_{\substack{i\\\ev{\Pi_{i}}=0}}\ev{A\Pi_{i}A}+\sum_{\substack{\text{degenerate\ }i\\\ev{\Pi_{i}}\neq0}}\ev*{A\Pi_{i}^{\psi^\perp}\!\! A}, 
    \label{eq: norm_rep}
\end{align}
where 
\begin{align}
    \Pi_{i}^{\psi^\perp}:=\Pi_{i}-\frac{\op{\Pi_{i}\psi}}{\ev{\Pi_{i}}}.
    \label{eq: ppjop_def}
\end{align}
We then recognize that there are two possibilities for $E(A, B) \ne 0$.  One of them is the case where the state $\ket{\psi}$ happens to be a zero eigenstate $\ev{\Pi_{i}}=0$ of the operator $\Pi_{i}$ for some $i$ and further the condition $\ev{A\Pi_{i}A} \ne 0$ holds.   The other, more generic, case occurs when the state $\ket{\psi}$ fulfills $\ev{\Pi_{i}} \ne 0$ together with $\ev*{A\Pi_{i}^{\psi^\perp}\!\! A} \ne 0$ for some degenerate ${i}$. 
In passing, we recall again that the quantity \eqref{eq: norm_subt} represents the discord  \cite{Lee_Tsutsui_1,Lee_Tsutsui_2,Lee_Tsutsui_3} between the proxy operator for $A$ by the measurement of $B$, and hence our examination of whether it vanishes or not is equivalent to whether the discord disappears or not.

%%%%%%%%%%%%%%%%%%
\begin{figure*}[t]
    \centering
    \begin{tabular}{ccc}
        \subfigure[]{
            \includegraphics[scale=0.44]{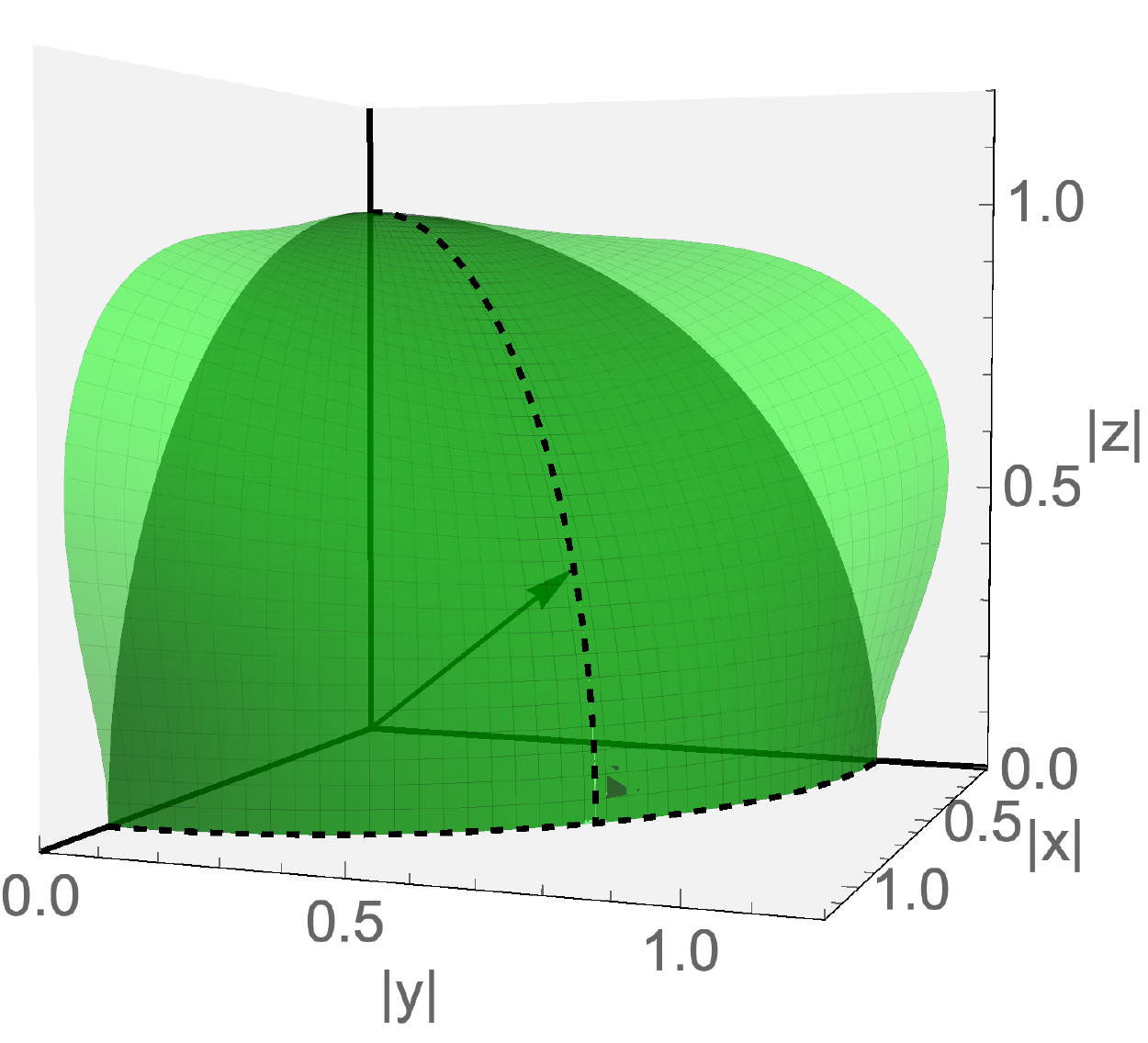}
        } &
        \subfigure[]{
            \includegraphics[scale=0.44]{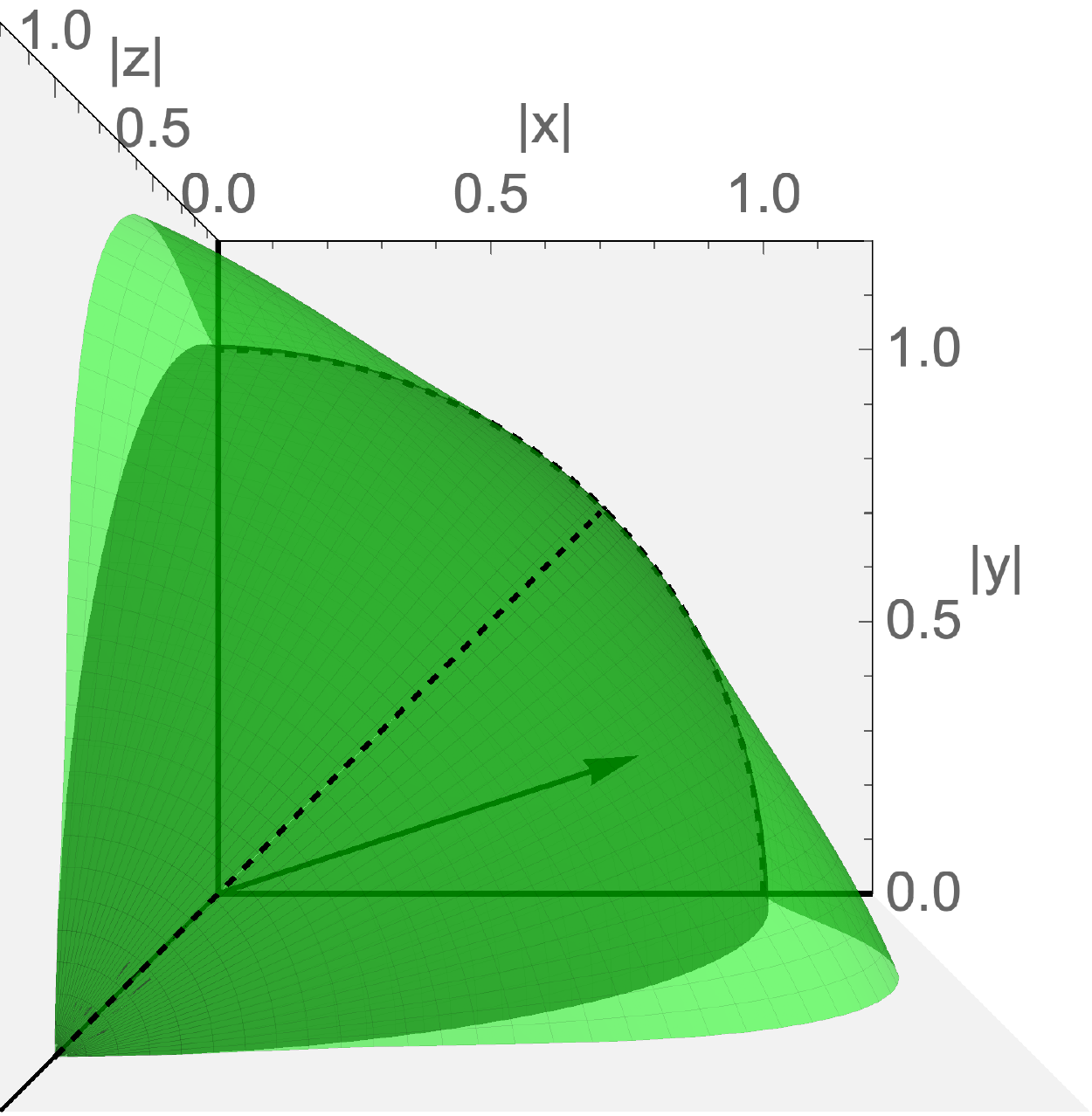}
        } &
        \subfigure[]{
            \includegraphics[scale=0.44]{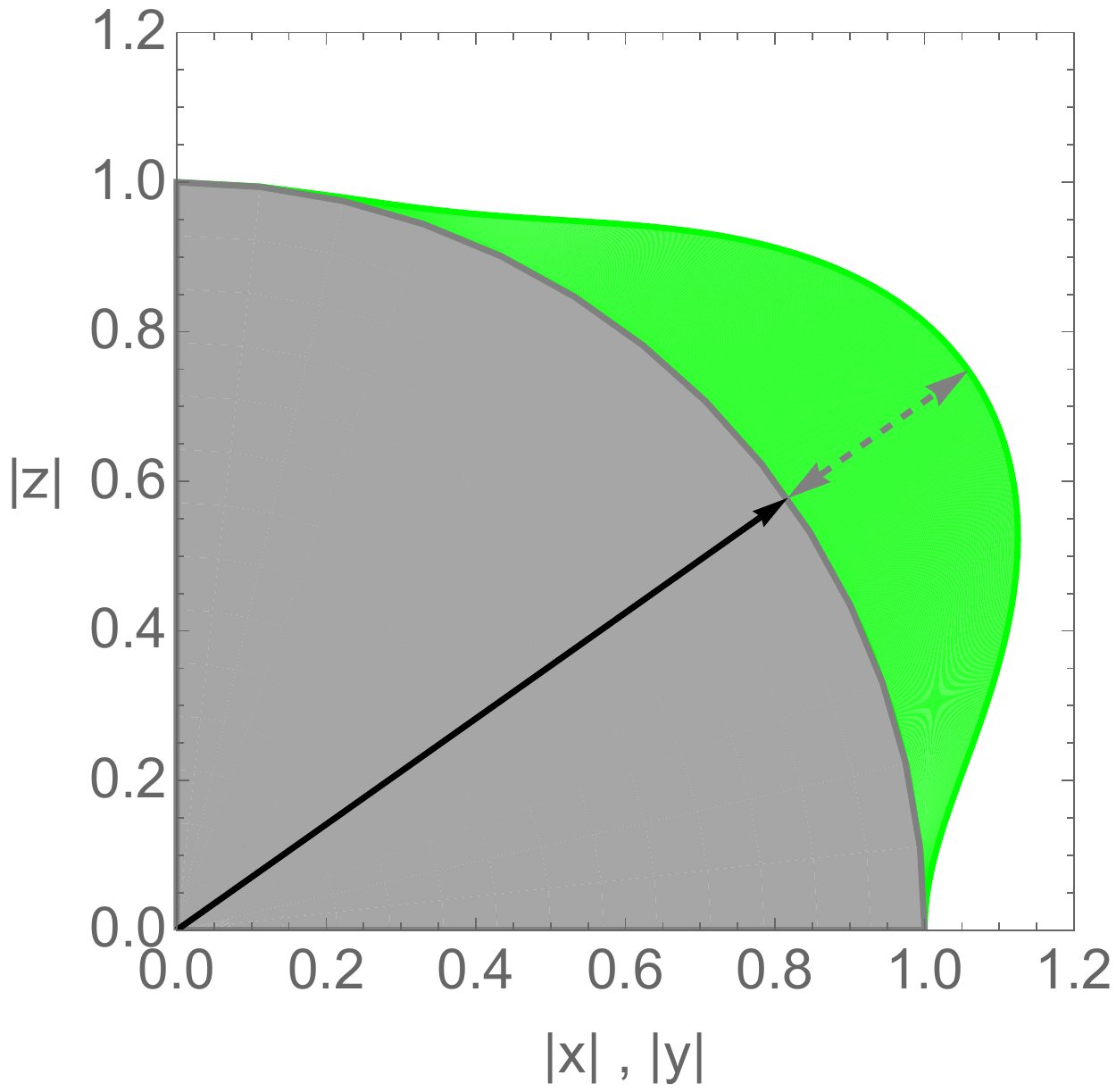}
        }
    \end{tabular}
    \caption{The extra term $E(A, B)$ evaluated in \eqref{eq: add_term} in the spin-1 example \eqref{eq: spex}. 
    The symbols $\abs{x}$, $\abs{y}$, and $\abs{z}$ correspond to each component of the state $\psi$ in \eqref{eq: spex}.  The states lie on the surface of the dark gray sphere with radius 1.  The radial distance between the inner sphere and the outer lilght green membrane (the length of the double-headed arrow in (c)) represents the value of the extra term $E(A, B)$ evaluated at the state $\psi$ specified by the arrow on the inner sphere.  Panel (c) shows the situation on the dissected $\abs{z}$-$\abs{x}$ and $\abs{z}$-$\abs{y}$ planes, where $E(A, B)$ is maximized when $\abs{z}=1/\sqrt{3}$ and $\abs{x}=0$ or $\abs{y}=0$.  Meanwhile, $E(A, B)$ vanishes when $\abs{x}=\abs{y}$ or $\abs{z}=0$ as illustrated by two dashed lines in (a) and (b).
    }
    \label{fig: spex}
\end{figure*}
%%%%%%%%%%%%%%%%%%

To confirm the first property (a) announced above, let us exclude from our consideration the special case where $B$ is proportional to the identity operator $I$, since in that case we have $\norm{\Delta B}^{2}=0$ and hence $E(A, B) = 0$ trivially.  Obviously, the cases $\dim\mathcal{H}=1$ and $\dim\mathcal{H}=2$ with a degenerate $B$
fall into this trivial category.  
For the case $\dim\mathcal{H}=2$ with a non-degenerate $B$, we find from \eqref{eq: norm_rep} that $\norm{A-A_{\rm w}(B)}^2\ne 0$ may be possible only when the state $\ket{\psi}$ fulfills either $\ev*{\Pi_{1}}=0$ or $\ev*{\Pi_{2}}=0$.  If $\ev*{\Pi_{1}}=0$, then one must have $\ket{\psi} = \ket{b_2}$ (up to a phase) for which 
$\norm{\Delta B}^{2}=\ev{B^{2}}-\ev{B}^{2} = 0$ again, putting $E(A, B) = 0$.  The other choice $\ev*{\Pi_{2}}=0$ does not alter the situation either, 
showing the first property (a) as we promised.

Next, to confirm the second property (b), we recall the fact that the inequality in \eqref{eq: tighter} is 
derived by applying the decomposed Schr{\"o}dinger inequality, \eqref{eq: cov} and \eqref{eq: KRlike}, in the decomposition 
\eqref{dadecop} obtained from the second term in the r.h.s.\ of \eqref{eq: da_pyth} while keeping the first term as it is. 
This suggests that the condition of equality remains also the same, namely, 
the lower bound of our inequality holds iff both of the conditions \eqref{eq: hold_cov} and \eqref{eq: hold_KRlike} are satisfied.

 Now, suppose that $B$ has only two distinct eigenvalues, $b_1$ and $b_2$.  For nontrivial situations $\norm{\Delta B}^{2} \ne 0$, we may assume $\ev{\Pi_{i}}\neq 0$ for $i = 1, 2$, since if $\ev{\Pi_{1}}=0$, say, then $\ev{\Pi_{2}}=1-\ev{\Pi_{1}}=1$ on account of $\Pi_{1} + \Pi_{2} = I$.   It follows that 
\begin{align}
    \norm{\Delta B}^{2} = b_2^2 \ev{\Pi_{2}}-b_2^2\ev{\Pi_{2}}^{2}=0,
\end{align}
showing that we have ended up with the trivial situation.   Once $\ev{\Pi_{i}}\neq 0$ is assured, then we may observe that
\begin{align}
    A_{\rm w}(B)-\ev{A} &=\sum_{i=1}^{2}\qty\Bigg(\frac{\ev{\Pi_{i}A}}{\ev{\Pi_{i}}}-\ev{\sum_{j=1}^{2}\Pi_{j}A})\Pi_{i} \nonumber\\
    &=\qty[\frac{\ev{\Pi_{1}A}}{\ev{\Pi_{1}}}(1-\ev{\Pi_{1}})-\ev{\Pi_{2}A}]\Pi_{1} \nonumber\\
    &\hphantom{=}+\qty[\frac{\ev{\Pi_{2}A}}{\ev{\Pi_{2}}}(1-\ev{\Pi_{2}})-\ev{\Pi_{1}A}]\Pi_{2} \nonumber\\
    &=(A_{\rm w}(b_{1})-A_{\rm w}(b_{2}))\,\qty[\ev{\Pi_{2}}\Pi_{1}-\ev{\Pi_{1}}\Pi_{2}], 
\end{align}  
and also that
\begin{align}  
    \Delta B&=\sum_{i=1}^{2}\qty\bigg(b_{i}-\sum_{j=1}^{2}b_{i}\ev{\Pi_{j}})\Pi_{i} \nonumber\\
    &=\qty[b_{1}(1-\ev{\Pi_{1}})-b_{2}\ev{\Pi_{2}}]\Pi_{1} \nonumber\\
    &\hphantom{=}+\qty[b_{2}(1-\ev{\Pi_{2}})-b_{1}\ev{\Pi_{1}}]\Pi_{2}\nonumber\\
    &=(b_{1}-b_{2})\,\qty[\ev{\Pi_{2}}\Pi_{1}-\ev{\Pi_{1}}\Pi_{2}],
\end{align}
to conclude
\begin{align}
    A_{\rm w}(B)-\expval{A} &=\frac{A_{\rm w}(b_{1})-A_{\rm w}(b_{2})}{b_{1}-b_{2}}\Delta B.
    \label{eq: d2_propto}
\end{align}
Operation of the real and imaginary parts of the both sides of \eqref{eq: d2_propto} on the state $\ket{\psi}$ gives \eqref{eq: hold_cov} and \eqref{eq: hold_KRlike}, respectively, with $\lambda$ and $\mu$ determined accordingly.  We thus see the second property (b) that the condition for attaining the lower bound of our inequality is satisfied if $B$ has only two distinct eigenvalues.  Note that this is valid for arbitrary $\psi$ and $A$.  

In the rest of this section, we will present two examples which exhibit these two properties. The first example is a simple spin-1 system with 
$\dim\mathcal{H}=3$.  Let $\sigma_{1} = \begin{pmatrix}
0 & 1 \\
1 & 0 \\
\end{pmatrix}$ and $\sigma_{2} = \begin{pmatrix}
0 & -i \\
i & 0 \\
\end{pmatrix}$ be the two Pauli matrices.  We choose then the two observables $A$, $B$ and the family of states
$\ket{\psi}$ as
\begin{align}
    A=\begin{pmatrix}
    \,\sigma_{1}\, & \rvline & \begin{matrix} \  \\\  \end{matrix} \\\hline
    \begin{matrix} & \end{matrix} &\rvline & 1
    \end{pmatrix},\ 
    B=\begin{pmatrix}
    \,\sigma_{2}\, & \rvline & \begin{matrix} \  \\\  \end{matrix} \\\hline
    \begin{matrix} & \end{matrix} &\rvline & 1
    \end{pmatrix},\ 
    \ket{\psi}=\frac{1}{\sqrt{N}}\begin{pmatrix}x\\y\\z\end{pmatrix},
    \label{eq: spex}
\end{align}
where $x,y,z\in\mathbb{C}$ and $N:=\abs{x}^2+\abs{y}^2+\abs{z}^2$.
With these, we can evaluate the extra term $E(A, B)$ in \eqref{eq: lastterm} explicitly as 
\begin{align}
    E(A, B)  = &\frac{2 \abs{z}^2}{N^3}(N-\abs{z}^{2}-2\abs{x}\abs{y}\cos\theta) \nonumber\\
    &\times (N-\abs{z}^{2}+2\abs{x}\abs{y}\sin\theta),
    \label{eq: add_term}
\end{align}
where $\theta:=\arg(x)-\arg(y)$ is the difference between the arguments of $x$ and $y$ (see Appendix \ref{app: s1_ex} for the derivation). Since $E(A, B)$ in \eqref{eq: add_term} depends only on $\theta$ and the absolute values of $x$, $y$, and $z$, we illustrate how it varies with respect to these parameters in Fig.~\ref{fig: spex}, assuming $N=1$, $\theta=0$ for simplicity. 
The fact that $E(A, B)$ is non-vanishing in this case ensures that our inequality is definitely tighter than the Sch\"rodinger inequality here. 
Furthermore, since $B$ in \eqref{eq: spex} has two distinct eigenvalues $\pm1$, the r.h.s.\ of our inequality \eqref{eq: tighter} is actually equals to the l.h.s., that is, the lower bound is attained as announced by the property (b).

%%%%%%%%%%%%%%%%%%
\begin{figure}[t]
    \centering
    \includegraphics[scale=0.52]{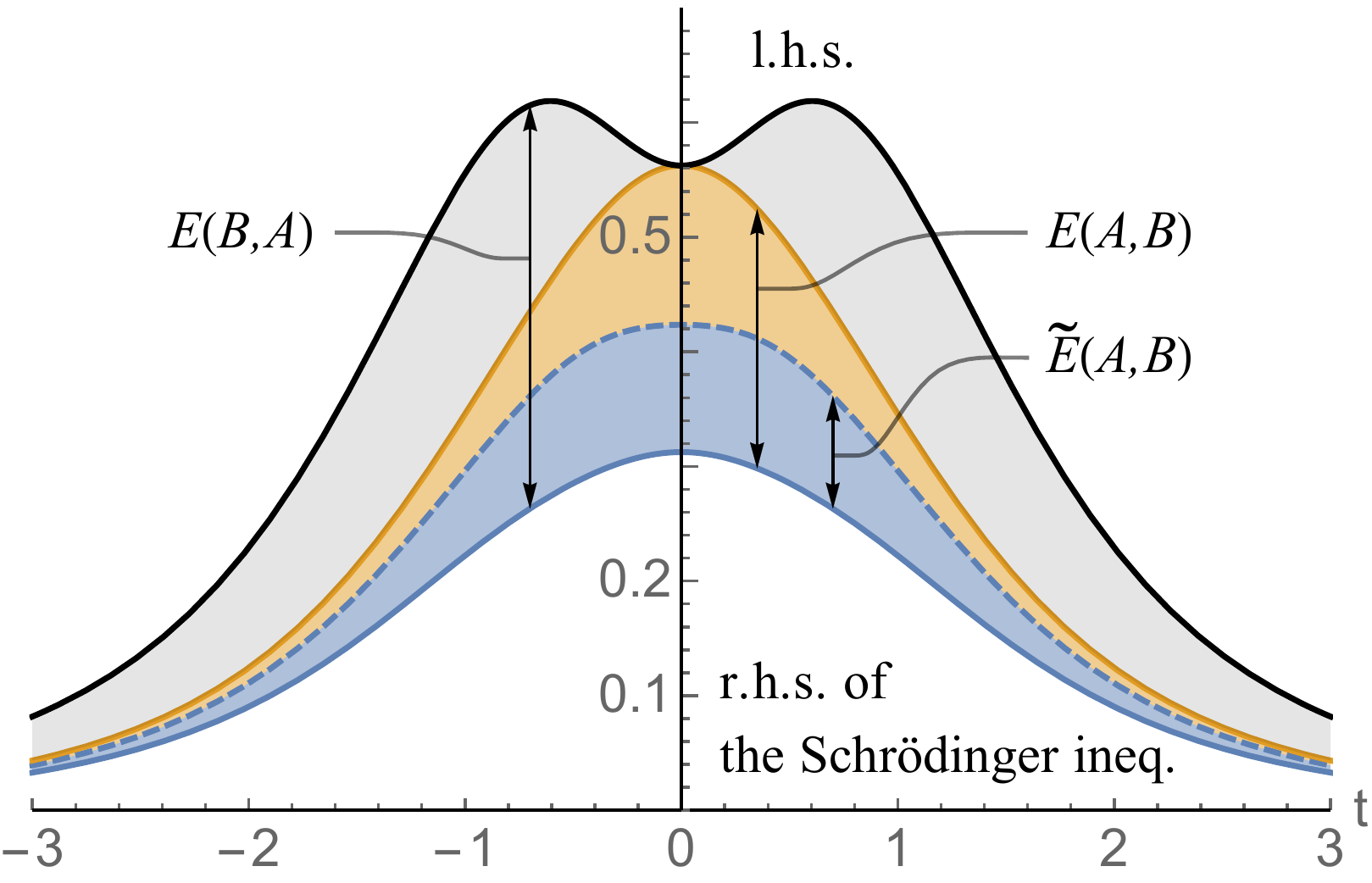}
    \caption{Contributions of the various extra terms to fill the gap between the l.h.s.\ and the r.h.s.\ of our inequality \eqref{eq: tighter}
evaluated for the spin-3/2 example \eqref{eq: wvtd4} as functions of $t$. 
The lowermost solid line represents the contribution of the r.h.s.\ of the Schr{\"o}dinger inequality \eqref{eq: Schr}, namely, the sum of the first and the second terms in the r.h.s.\ of \eqref{eq: tighter}.  The uppermost solid line 
represents the l.h.s., which is just the product of the variances $\norm{\Delta A}^{2} \norm{\Delta B}^{2}$.}
\label{fig: wvtd4}
\end{figure}
%%%%%%%%%%%%%%%%%%

Next we consider a little more complicated example of a spin-3/2 system with $\dim\mathcal{H}=4$.   For the observables and states, 
let us choose 
\begin{align}
    A=\begin{pmatrix}
    \,\sigma_{1}\, & \rvline & \begin{matrix} & \\ & \end{matrix} \\\hline
    \begin{matrix} & \\ & \end{matrix} &\rvline & \begin{matrix} 1 & \\ &-1 \end{matrix}
    \end{pmatrix},\,
    B=\begin{pmatrix}
    \,\sigma_{2}\, & \rvline & \begin{matrix} & \\ & \end{matrix} \\\hline
    \begin{matrix} & \\ & \end{matrix} &\rvline & \begin{matrix} 1 & \\ & 0 \end{matrix}
    \end{pmatrix},\,
    \ket{\psi}=\frac{1}{\sqrt{N'}}\begin{pmatrix}1\\0\\1\\t\end{pmatrix}.
    \label{eq: wvtd4}
\end{align}
with $t\in\mathbb{R}$ and $N':=2+\abs{t}^2$.  
Fig.~\ref{fig: wvtd4} illustrates the various relevant terms related to our inequality, 
which shows that the extra terms are indeed non-vanishing.  It also shows that the lower bound is not attained with either the extra terms
$E(A, B)$ and $\widetilde{E}(A, B)$, while it is 
attained with $E(B, A)$ (which is equivalent to $E_{\text{max}}(A, B)$ in this case).   
The reason for the latter is that, since $A$ has only two distinct eigenvalues $\pm1$, 
the property (b) ensures that $E(B, A)$, instead of $E(A, B)$, furnishes the equality.  
In short, we have seen here an example where 
the Sch\"rodinger inequality never attains its lower bound while our improved inequality does with the proper choice of the extra term.

%%%%%%%%%%%%%%%%%%%%%%%
\section{Remarks on the Continuous Case}
\label{sec: cont}
%%%%%%%%%%%%%%%%%%%%%%%

In the previous two sections, we have reviewed the decomposed form \cite{Lee_Tsutsui_1, Lee_Tsutsui_2} of the Schr{\"o}dinger inequality \eqref{eq: cov}, \eqref{eq: KRlike} and the tighter inequality \eqref{eq: tighter} as well as the associated two conditions for the minimum uncertainty \eqref{eq: hold_cov}, \eqref{eq: hold_KRlike}. 
We now consider the case of continuous observables by taking up the familiar example of a particle moving on a line and using the momentum and the position as the two observables, $A=\hat{p},\ B=\hat{x}$.  In order to treat $\hat{p}$ and $\hat{x}$, we pretend in this paper that all the important properties we have recapitulated so far remain valid even for continuous observables, once we formally \lq replace\rq\ the summation over the indices $i$ by integration over the spectrum; as we have mentioned earlier, this is in fact guaranteed for the general case with full rigour \cite{Lee_Tsutsui_2,Lee_Tsutsui_3}. 

Let us first inspect the states fulfilling the conditions, \eqref{eq: hold_cov} and \eqref{eq: hold_KRlike}, separately. 
Multiplying the both sides of \eqref{eq: hold_cov} by
$\bra{x}$ from left, we have
\begin{align}
    \Im\frac{\psi^{\prime}(x)}{\psi(x)} = \frac{\lambda}{\hbar}(x-\expval{x})+\frac{\expval{p}}{\hbar},
    \label{eq: difeq_phs}
\end{align}
where $\psi(x):=\braket{x}{\psi}$ is the wave function. The general solution of \eqref{eq: difeq_phs} is then found as
\begin{align}
    \psi(x) = C_{1}\exp[i\qty(\frac{\lambda}{2\hbar}(x-\expval{x})^2+\frac{\expval{p}}{\hbar}x)] e^{a(x)},
    \label{eq: min_phs}
\end{align}
with a constant $C_{1}$ and an arbitrary real function $a(x)$.  We thus learn that the condition \eqref{eq: hold_cov} determines the state in the phase part.  
Similarly, from \eqref{eq: hold_KRlike} we have
\begin{align}
    \Re\frac{\psi^{\prime}(x)}{\psi(x)} = -\frac{\mu}{\hbar}(x-\expval{x}),
\end{align}
whose general solution is
\begin{align}
    \psi(x) = C_{2}\exp[-\frac{\mu}{2\hbar}(x-\expval{x})^2] e^{i\theta(x)},
    \label{eq: min_dis}
\end{align}
with a constant $C_{2}$ and an arbitrary real function $\theta(x)$. 
This shows that the condition \eqref{eq: hold_KRlike} determines the state in the modulus part.

Since the minimum uncertainty states of the Schr{\"o}dinger inequality must be of the forms \eqref{eq: min_phs} and \eqref{eq: min_dis}, simultaneously, they take the form,
\begin{align}
    \psi(x) = C&\exp[i\qty(\frac{\lambda}{2\hbar}(x-\ev{x})^2+\frac{\ev{p}}{\hbar}x)] \nonumber\\
    \times &\exp[-\frac{\mu}{2\hbar}(x-\ev{x})^2], 
    \label{eq: min_Schr}
\end{align}
where $C$ is a constant.

Of course, we can obtain the solution \eqref{eq: min_Schr} directly from the condition \eqref{eq: hold_Schr} for the minimum uncertainty states of the original Schr{\"o}dinger inequality \eqref{eq: Schr}.  Indeed, by operating $\bra{x}$ from the left on the condition \eqref{eq: hold_Schr}, we obtain
\begin{align}
    \qty(\frac{\hbar}{i}\dv{x}-\ev{p})\psi(x)=(\lambda+i\mu)(x-\ev{x})\psi(x),
    \label{eq: diff_px}
\end{align}
which admits the solution $\psi(x)=\exp[S(x)]$ with 
\begin{align}
    S(x)=i\qty(\frac{\lambda}{2\hbar}(x-\ev{x})^2+\frac{\ev{p}}{\hbar}x)-\frac{\mu}{2\hbar}(x-\ev{x})^2+C^{\prime},
\end{align}
where $C^{\prime}$ is a constant, reproducing the solution \eqref{eq: min_Schr}.  The benefit of using the decomposed conditions 
lies in the fact that it allows us to see how the solution arises from each of the conditions, which in turn arise from the two CS inequalities used to derive the Schr\"odinger inequality \eqref{eq: Schr}.

Finally, we briefly touch upon our inequality \eqref{eq: tighter} applied to the present continuous case.
From \eqref{eq: norm_rep}, the norm $\norm{A-A_{\rm w}(B)}^{2}$ appearing in the first term of the r.h.s.\ of \eqref{eq: tighter} reads 
\begin{align}
    \norm{\hat{p}-\hat{p}_{\rm w}(\hat{x})}^{2}&=\int_{\psi(x)=0}\mel{\psi}{\hat{p}}{x}\mel{x}{\hat{p}}{\psi}dx \nonumber\\
    &=\hbar^{2} \int_{\psi(x)=0} \frac{\mathrm{d} \psi^{*}(x)}{\mathrm{d} x} \frac{\mathrm{d} \psi(x)}{\mathrm{d} x} dx. 
    \label{eq: pwxnorm}
\end{align}
Normally, we may encounter $\psi(x) = 0$ either at discrete points (as in the case of excited states of a harmonic oscillator) or continuously for a certain range (as in the case of a localized wave packet). 
However, in the former case, such points are measure zero and hence 
the integration \eqref{eq: pwxnorm} yields $E(\hat{p}, \hat{x}) =0$.  Likewise, 
in the latter case we have $\psi^{\prime}(x)=0$ for the range, yielding again $E(\hat{p}, \hat{x}) =0$, as long as we are dealing with smooth wave functions.  Therefore, in order for us to have $E(\hat{p}, \hat{x}) \ne 0$, we must deal with wave functions which are singular at some set of points, although it is not clear if it is admitted by a physically interesting system.

%%%%%%%%%%%%%%%%%%%%%%%
\section{conclusion and discussions}
\label{sec: condis}
%%%%%%%%%%%%%%%%%%%%%%%

In this paper, we have analyzed in detail the structure of the Schr{\"o}dinger inequality \cite{Schroedinger} which underlies the Kennard-Robertson inequality \cite{Kennard, Robertson}, which is often regarded as the standard form of the uncertainty relation for the product of variances of two arbitrary observables $A$ and $B$.  
Our basic tool for the analysis is the weak-value operator $A_{\rm w}(B)$ introduced in our previous works \cite{Lee_Tsutsui_1,Lee_Tsutsui_2,Lee_Tsutsui_3} as the optimal operator for estimating $A$ by measuring $B$, which was named after Aharonov's weak value \cite{ABL,AAV}.  This tool provided us with a useful geometrical picture of the operator space acting on the Hilbert space of the quantum system, thereby yielding the decomposition of the Schr{\"o}dinger inequality into two components \cite{Lee_Tsutsui_1,Lee_Tsutsui_2} based on the complex (non-selfadjoint) nature of the weak-value operator $A_{\rm w}(B)$;  roughly speaking, the Schr{\"o}dinger inequality had been decomposed into the real and the imaginary parts accordingly.

The decomposition allows us to find, for instance, how the minimal uncertainty states take their shape with respect to the real and imaginary components, separately.  This is exemplified in the case of $A$ and $B$ being position and momentum operators, where the phase and the modulus parts of the familiar minimal uncertainty states, {\it i.e.}, the coherent states, are seen to be determined, respectively, by the conditions obtained from the two decomposed components of the Schr{\"o}dinger inequality.  

Another, perhaps more significant, benefit of using the weak-value operator $A_{\rm w}(B)$ is that it leads us to a novel inequality \eqref{eq: tighter} that reveals an additional contribution $E(A, B)$ to the lower bound of the Schr{\"o}dinger inequality; although one may argue that the additional term $E(A, B)$ vanishes in many cases, it is not difficult to find examples for which it does not.

Of particular interest is thus to know precisely when the extra term $E(A, B)$ becomes non-vanishing, rendering our inequality genuinely different from the Schr{\"o}dinger inequality.  From the fact that $E(A, B)$ is directly proportional to the distance between $A_{\rm w}(B)$ and $A$ in the operator space, a straightforward strategy would be to to look for the condition on which the distance becomes non-vanishing; our previous works imply that this could be attained when, for example, the given state happens to have a zero inner-product with an eigenstate of $B$, or when $B$ is degenerate for some of its eigenvalues.  These cases may occur when the state is prepared to be orthogonal to one or more of the eigenstates of $B$, or when the degeneracy of $B$ is ensured from the symmetry of the system.  Interestingly, our inequality becomes an equality if the operator $B$ has only two distinct eigenvalues, which indicates that for lower dimensional systems the lower bound is pretty tight.  To illustrate these findings, we have furnished elementary examples of spin-1 and 3/2 models in which $E(A, B)$ is non-vanishing.  Our quantitative assessment of the terms appearing in the supplemented Schr{\"o}dinger inequality shows that different types of extra terms which are allowed yields distinct contributions for the lower bound.

It should also be noted that the presence of the extra term in the lower bound can alter the conditions for the minimal uncertainty states.  Accordingly, we need to examine if $E(A, B)$ vanishes in addition to confirming if the equality conditions are met.  In the familiar example of position and momentum, however, it is implied that the extra term vanishes identically as long as smooth wave functions are concerned, assuring that the minimal uncertainty states still remain the same.  

Our inequality \eqref{eq: tighter} with the extra term \eqref{eq: lastterm} permits the interpretation that the source of the lower bound for the product of variances is a combination of quantum noncommutativity, classical covariance, and the measurement discord \cite{Lee_Tsutsui_1,Lee_Tsutsui_2,Lee_Tsutsui_3} between the two observables.  The discord depends on the choice of measured operator, and one can put it in the symmetrized form \eqref{eq: tighter_indp} by adopting the discord from the opposite choice of measured operator, eliminating the variance altogether from the lower bound.  

In all aspects of our analysis, we have seen that the weak-value operator plays a crucial role, exhibiting perhaps an unexpected importance of the notion of the weak value in quantum uncertainty.  We hope that, besides finding further applications in actual physical systems in estimating uncertainty closely, our inequality presented here serves as a testing ground for examining the significance of the weak value in a broader context of quantum foundations.

\begin{acknowledgments}
This work was supported by JSPS Grant-in-Aid for Scientific Research (KAKENHI), Grant Numbers JP18K13468 and JP18H03466.
\end{acknowledgments}

%%%%%%%%%%%%%%%%%%%%%%%%%%%%%%%
\appendix
%%%%%%%%%%%%%%%%%%%%%%%%%%%%%%%

%%%%%%%%%%%%%%%%%%%%%%%%%%%%%%%
\section{Supplement to the spin-1 example}
\label{app: s1_ex}
%%%%%%%%%%%%%%%%%%%%%%%%%%%%%%%

Here we provide some supplemental material for the spin-1 example \eqref{eq: spex}. 
We will derive \eqref{eq: add_term} and confirm the equality in \eqref{eq: tighter} mentioned in Sec.~\ref{sec: cond}.

Let us label the eigenvalues of $B$ as $b_{1}=-1$, $b_{2}=1$.  The projectors associated with these eigenvalues then read
\begin{align}
    \Pi_{1}=\frac{1}{2}\begin{pmatrix}
        1 & i & 0 \\
       -i & 1 & 0 \\
        0 & 0 & 0 \\
    \end{pmatrix}, \quad
    \Pi_{2}=\frac{1}{2}\begin{pmatrix}
        1 &-i & 0 \\
        i & 1 & 0 \\
        0 & 0 & 2 \\
    \end{pmatrix}.
\end{align}
By introducing the abbreviation,
\begin{align}
    c := 2\abs{x}\abs{y}\cos\theta ,\quad s := 2\abs{x}\abs{y}\sin\theta,
\end{align}
the variances of the $A$ and $B$ are written as
\begin{align}
    \norm{\Delta A}^{2} &= \frac{N^2-(\abs{z}^2 + c)^2}{N^2},
    \label{eq: var_A}\\
    \norm{\Delta B}^{2} &= \frac{N^2-(\abs{z}^2 - s)^2}{N^2}.
    \label{eq: var_B}
\end{align}

To derive \eqref{eq: add_term}, we first focus on the second line of \eqref{eq: wvonorm},
\begin{align}
    &\norm{A_{\rm w}(B)}^{2} \nonumber\\
    &=\norm\bigg{\sum_{\substack{i\\\ev{\Pi_{i}}\neq0}}\frac{\ev{\Pi_{i}A}}{\ev{\Pi_{i}}}\Pi_{i}}^{2} \nonumber\\
    &=\norm{\frac{c+i(\abs{x}^{2}-\abs{y}^{2})}{N-\abs{z}^{2}+s}\Pi_{1} + \frac{c+2\abs{z}^{2}-i(\abs{x}^{2}-\abs{y}^{2})}{N+\abs{z}^{2}-s}\Pi_{2}}^{2} \nonumber\\
    &=\frac{N(N-\abs{z}^{2}-s)+2\abs{z}^{2}(\abs{z}^{2}+c)}{N(N+\abs{z}^{2}-s)}.
\end{align}
Thus, from \eqref{eq: norm_subt} we have
\begin{align}
    \norm{A-A_{\rm w}(B)}^{2} &= 1 - \norm{A_{\rm w}(B)}^{2} = \frac{2 \abs{z}^2}{N}\cdot\frac{N-\abs{z}^{2}-c}{N+\abs{z}^{2}-s}.
    \label{eq: norm_spex}
\end{align}
The product of \eqref{eq: var_B} and \eqref{eq: norm_spex} leads to the extra term \eqref{eq: add_term} to form the last term in the r.h.s.\ of \eqref{eq: tighter} as expected.

Next, we confirm the correspondence between the l.h.s.\ and r.h.s.\ of \eqref{eq: tighter}. The second and third terms in the r.h.s.\ are
\begin{align}
    \abs{\frac{1}{2}\ev{\acomm{A}{B}}-\ev{A}\ev{B}}^{2} &= \qty[\frac{\abs{z}^2 N-(\abs{z}^{2} + c) (\abs{z}^{2}-s)}{N^2}]^{2}
    \label{eq: spex_cov} \\
    \abs{\frac{1}{2}\ev{\comm{A}{B}}}^{2} &= \qty[\frac{\abs{x}^2-\abs{y}^2}{N}]^2,
    \label{eq: spex_comm}
\end{align}
respectively. Noting that
\begin{align}
    (\abs{x}^2-\abs{y}^2)^2 = (N-\abs{z}^2)^2 - (c^2 + s^2)
\end{align}
holds in \eqref{eq: spex_comm}, we see that the sum of the three terms in the r.h.s., \eqref{eq: add_term}, \eqref{eq: spex_cov}, and \eqref{eq: spex_comm} yields
\begin{align}
    1 - \frac{(\abs{z}^{2} + c)^{2} + (\abs{z}^{2}-s)^{2}}{N^2}+\frac{(\abs{z}^{2} + c)^{2} (\abs{z}^{2}-s)^{2}}{N^4},
\end{align}
which agrees with the  product of \eqref{eq: var_A} and \eqref{eq: var_B} as promised.

%%%%%%%%%%%%%%%%%%%%%%%
\section{Geometrical interpretation on the decomposed terms}
\label{app: geom_int}
%%%%%%%%%%%%%%%%%%%%%%%

\begin{figure*}[t]
    \centering
    \includegraphics[scale=0.24]{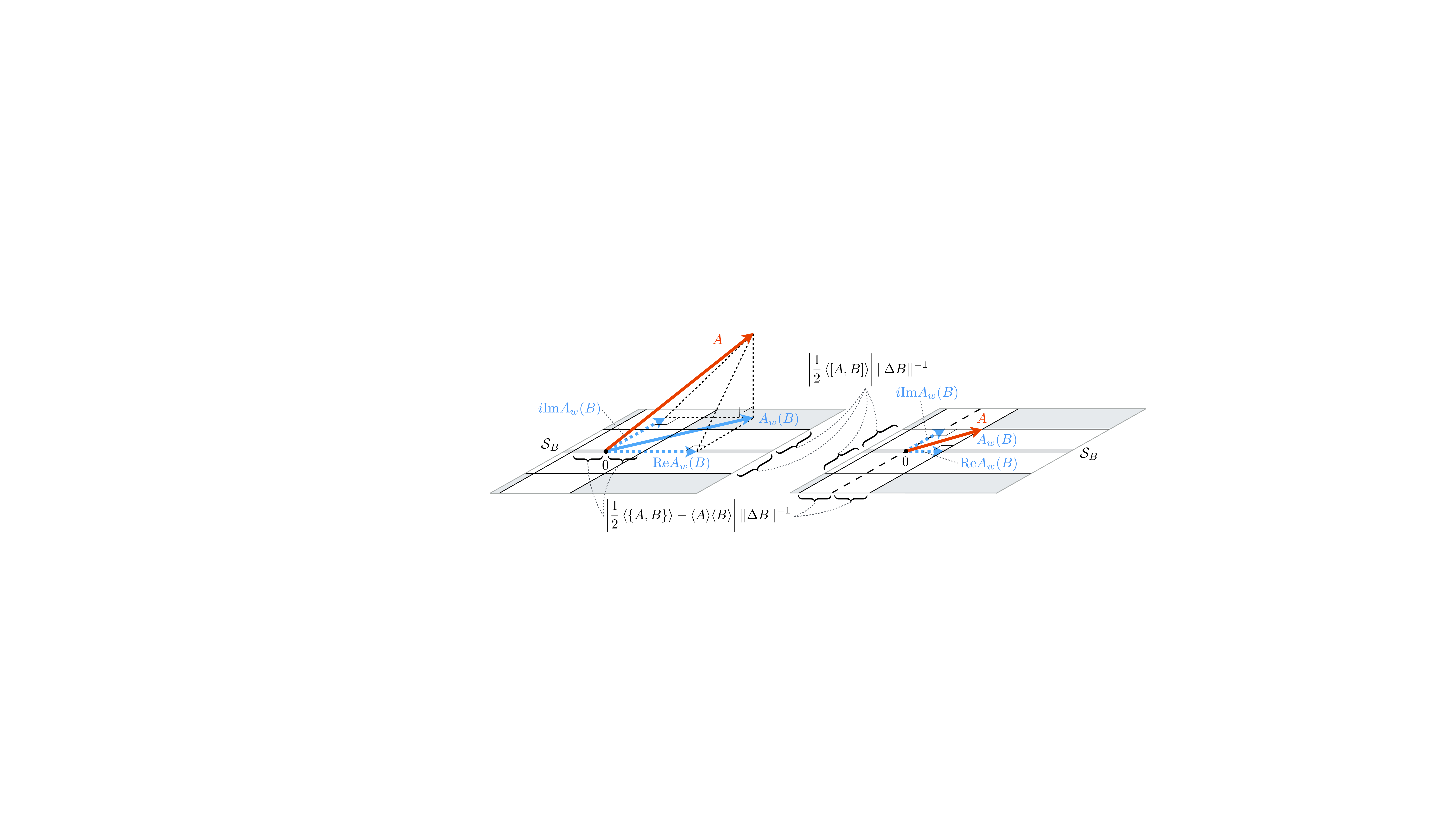}
    \caption{Geometric relations among the operators involved in the uncertainty relations of variances. For simplicity, $\ev{A}=0$ is assumed, which only offsets the origin. 
    Each plane lying in the left and right represents ${\cal L}_B$, 
    and each light gray centerline represents ${\cal S}_B$. 
    The black solid lines represent the lower bound given by the decomposed form of the Schr{\"o}dinger inequality, \eqref{eq: cov} and \eqref{eq: KRlike}; thus $A_{\rm w}(B)$ is bounded in the shaded region. Note that all of the lengths appear in this figure depend on the state $\psi$.
    The left depicts a generic state, while the right depicts the case where $\norm{A-A_{\rm w}(B)}=0$ holds and $\psi$ satisfies the lower bound of the Schr{\"o}dinger inequality \eqref{eq: Schr}.
    }
    \label{fig: ineq_geom}
\end{figure*}

Finally, we offer below a detailed visual comparison among the various terms of the Schr{\"o}dinger equality in view of the geometric description introduced in our previous works \cite{Lee_Tsutsui_1,Lee_Tsutsui_2}, which we have also reviewed in Sec.~\ref{sec: geom_drv}.

Dividing both sides of our inequality \eqref{eq: tighter} by $\norm{\Delta B}$, we can illustrate 
the operators involved in each of the terms as shown in Fig.~\ref{fig: ineq_geom}. 
Here, the position of $A_{\rm w}(B)$, which is projected from $A$ onto the plane ${\cal L}_B$, lies in the shaded regions on account of the decomposed components, \eqref{eq: cov} and \eqref{eq: KRlike}. 
When the state $\psi$ attains the lower bound,
$A_{\rm w}(B)$ resides at the intersections of the black solid lines in Fig.~\ref{fig: ineq_geom}.
The lower bound of the Schr{\"o}dinger inequality \eqref{eq: Schr}, on the other hand, can be attained only when $\norm{A-A_{\rm w}(B)}=0$ holds, namely $A$ corresponds to $A_{\rm w}(B)$ as shown in the right of Fig.~\ref{fig: ineq_geom}, at which point the two novel relations are equivalent to this inequality.

In addition, when the covariance term $\abs{\frac{1}{2}\ev{\acomm{A}{B}}-\ev{A}\ev{B}}$ is zero, then we end up with the KR inequality \eqref{eq: KR}. Now, in the right of Fig.~\ref{fig: ineq_geom}, the lines representing the lower bound \eqref{eq: cov} of the covariance term reduce to the dashed line, $\Re A_{\rm w}(B)$ reduces to zero, and $A$ (and $A_{\rm w}(B)$) corresponds to $i\Im A_{\rm w}(B)$.

%%%%%%%%%%%%%%%%%%%%%%%%%%%%%%%
\bibliography{Schroedinger_ref0807}
%%%%%%%%%%%%%%%%%%%%%%%%%%%%%%%

\end{document}